\documentclass[aps,prl,reprint,letterpaper,superscriptaddress,showpacs]{revtex4-1}
\usepackage{CJK}
\usepackage{keyval}%
\usepackage{graphicx}
\usepackage{dcolumn}
\usepackage{bm}
\usepackage{color}
\usepackage{amssymb}
\usepackage{amsmath}
\setkeys{Gin}{width=0.95\columnwidth,clip=true}

\newcommand{\ignore}[1]{}

\newcommand{\liro}{Li$_2$Ir$_{1-x}$Ru$_x$O$_3$}
\newcommand{\lrro}{Li$_2$Rh$_{1-x}$Ru$_x$O$_3$}
\newcommand{\lio}{Li$_2$IrO$_3$}
\newcommand{\lro}{Li$_2$RuO$_3$}

\newcommand{\siro}{Sr$_2$Ir$_{1-x}$Ru$_x$O$_4$}
\newcommand{\sio}{Sr$_2$IrO$_4$}

\begin{document}
\title{
Structural, Magnetic, and Electrical Properties of \liro}
\author{Hechang Lei }
\affiliation{Frontier Research Center and Materials and Structures Laboratory, Tokyo Institute of Technology, 4259 Nagatsuta, Midori, Yokohama 226-8503, Japan}
\author{Wei-Guo Yin} 
\email{wyin@bnl.gov}%
\affiliation{Condensed Matter Physics and Materials Science Department,
Brookhaven National Laboratory, Upton, New York 11973, USA}
\author{Zhicheng Zhong} 
\affiliation{Institute of Solid State Physics, Vienna University of Technology, A-1040 Vienna, Austria}
\author{Hideo Hosono}  
\email{hosono@msl.titech.ac.jp}
\affiliation{Frontier Research Center and Materials and Structures Laboratory, Tokyo Institute of Technology, 4259
Nagatsuta, Midori, Yokohama 226-8503, Japan}

\date{\today}

\begin{abstract}

The crystal structure, resistivity, and magnetic susceptibility of the \liro\ ($x$ = 0$-$1) polycrystals have been investigated. We found that the parent antiferromagnetic phase disappears for $x > 0.2$ and bond dimers appear in the averaged structure for $x>0.5$ and likely fluctuate for much smaller $x$. Unexpectedly, this system remains insulating for all the doping levels, in sharp contrast with the robust metallic state found in \siro.
These results demonstrate that the essential physics of the doped A$_2$IrO$_3$ (A = Li, Na) system deviates significantly from the common one-band $j_\mathrm{eff}=1/2$ spin-orbit model.
\end{abstract}


\maketitle

A variety of insulating iridium oxides with open Ir$^{4+}$ $5d$ shells, such as honeycomb A$_2$IrO$_3$ (A = Li and Na), square-lattice A$_2$IrO$_4$ (A = Sr and Ba), hyperkagome-lattice
Na$_4$Ir$_3$O$_8$, and pyrochlores R$_2$Ir$_2$O$_7$ (R = Y, Sm, Eu, and Lu), are a subject of recent intensive investigations. These iridates break the general expectation that open-shell $5d$ systems are wide-band weakly correlated metals, and are characterized as `spin-orbit Mott insulators' owing to the band-narrowing effect of strong spin-orbit coupling (SOC) on the Ir $5d$ orbitals \cite{Sr2IrO4:Kim_Science09,Sr2IrO4:Kim_08,Sr2IrO4:Moon}. It is beginning to explore what exotic phenomena can be developed from this novel type of Mott insulators \cite{A2Ir2O7:Pesin}.

One particularly interesting question is how the magnetic and electric properties of these iridates evolve upon charge doping---will superconductivity emerge \cite{PhysRevB.86.125105,PhysRevB.84.100402,Sr2IrO4:magnon:Kim,PhysRevLett.110.027002,PhysRevB.85.140510,PhysRevB.86.085145,PhysRevB.87.064508}? This was motivated by comparing the iridates to the layered cuprates in which high-temperature superconductivity develops when the `parent' antiferromagnetic (AFM) Mott insulating phase is suppressed by doping. Similarly, the layered iridates A$_2$IrO$_4$ (214) and A$_2$IrO$_3$ (213) exhibit long-range AFM ordered ground states as well \cite{Sr2IrO4:magnon:Kim,PhysRevB.82.064412,Na2IrO3:zigzag:Liu,Na2IrO3:zigzag:Choi,Na2IrO3:zigzag:Ye,Na2IrO3:Singh}. In addition, both the 214 and 213 iridates have been widely modeled as effective one-band total angular momentum $j_\mathrm{eff}=1/2$ Mott insulators \cite{Sr2IrO4:Kim_Science09,Sr2IrO4:Kim_08,Sr2IrO4:Moon,A2Ir2O7:Pesin,Ir:Jackeli}, comparable to the effective one-band spin $S=1/2$ Mott insulator modeling of the cuprates. Experimentally, Ru, La or K doping was found to systematically drive \sio\ to a robust metallic state, although superconductivity is not yet at reach  \cite{PhysRevB.86.125105,PhysRevB.84.100402}. As for A$_{2}$IrO$_{3}$, the deviation of the observed AFM state from the predicted Kitaev spin-liquid (KSL) state \cite{Ir:Jackeli} seems to be remedied by including the Heisenberg exchanges, as the calculated magnon and single-hole spectra agree with the experiments \cite{Na2IrO3:zigzag:Choi,PhysRevLett.109.266406,PhysRevLett.110.097204,PhysRevLett.111.037205,arXiv:1308.3373}. 
Based on the one-band $j_\mathrm{eff}=1/2$ Kitaev-Heisenberg model, it was further predicted that the superconductive ground state would emerge with hole doping \cite{PhysRevB.85.140510,PhysRevB.86.085145,PhysRevB.87.064508} but the experimental information is lacking.

In this Letter, we present experimental studies of \liro\ ($x$ = 0$-$1) polycrystals on crystal structure, resistivity, magnetic susceptibility. Since the two end members have nominal Ir$^{4+}$ $5d^5$ and Ru$^{4+}$ $4d^4$, respectively, Ru substitution for Ir is generally regarded as hole doping. The essential crystal structure is the honeycomb lattice of the Ir/Ru atoms [Fig.~1(a)].  In \lio\ all the Ir-Ir bond lengths are almost the same. Whereas, in \lro\ one third of the Ru-Ru bonds are significantly shortened below a metal-insulator transition at $540$~K, forming ordered dimers ascribed as molecular orbitals \cite{JPSJ.76.033705,JPSJ.78.094706,arXiv:1310.7810} or spin singlets \cite{PhysRevLett.100.147203}. It was anticipated that once the structural phase transition is suppressed, Li$_{2}$RuO$_{3}$ should be metallic \cite{JPSJ.78.094706}. As expected, we observed that the AFM order in \lio\ and the bond-length alternation in \lro\ are suppressed by doping the system away from the end members.

Surprisingly, we found that \liro\ remains insulating for all the doping levels. This sharp contrast between the 213 and 214 iridate systems demonstrates that the fundamental physics of the doped iridates depends strongly on the lattice structure. For the honeycomb lattice, we attribute the hole induced breakdown of the $j_\mathrm{eff}=1/2$ picture to a novel quasi-molecular-orbital Jahn-Teller instability, which reactivates the orbital degree of freedom (d.o.f.) and leads to the large effects of electron-phonon (EP) coupling which cooperates with electron-electron interaction to account for the persistent insulating character and bond dimerization.

Polycrystalline samples of \liro\ were synthesized using a solid-state reaction method as described previously \cite{Na2IrO3:Singh,JPSJ.76.033705}. RuO$_{2}$ was heated at 1000 K for 6~h and Li$_{2}$CO$_{3}$ was baked at 500 K for 5 h in air before use. Stoichiometric amounts of Li$_{2}$CO$_{3}$, RuO$_{2}$ and anhydrous IrO$_{2}$ were mixed, ground, and pelletized. Then, the pellets were placed in alumina crucible which was covered by lid and sintered at 975 ${{}^{\circ }}C$ for 24 h, followed by furnace cooling to room temperature. The resulting materials were mixed with 5 \% Li$_{2}$CO$_{3}$ in order to compensate the loss of Li$_{2}$CO$_{3}$ during heating treatment. The mixtures were reground, pelletized and sinter at 975~$^{\circ}C$ for 48~h. This step was repeated for several times until the final samples were pure \liro\ without trace of RuO$_{2}$. The structures of the samples were characterized by powder x-ray diffraction (XRD) using a Bruker diffractometer model D8 ADVANCE (reflection mode with Cu $K_{\alpha}$ radiation and transmission mode with Mo $K_{\alpha}$ radiation and capillary). Rietveld refinement of the XRD patterns was performed using the code TOPAS4 \cite{Topas}. Electrical transport with four-probe configuration and high-temperature magnetization measurements were carried out in Quantum Design PPMS-9. Low-temperature magnetization measurement was carried out in Quantum Design MPMS SQUID VSM.


\begin{figure}[tbp]
\centerline{\includegraphics[scale=0.16]{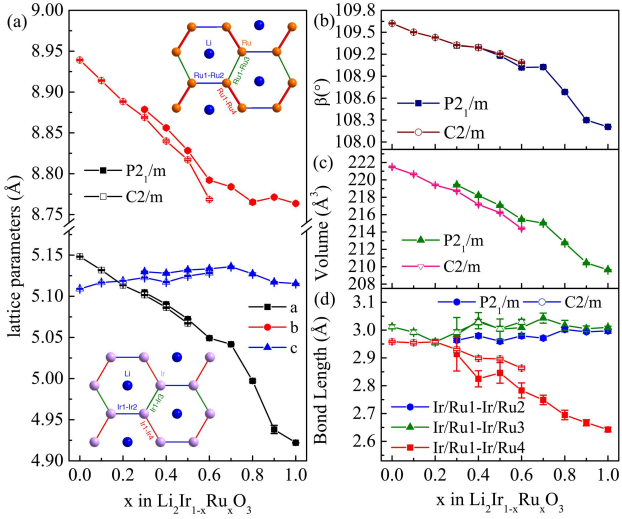}} \vspace*{-0.3cm}
\caption{(a) Lattice parameters, (b) the $\beta$ values, (c) the unit-cell volumes,  and (d) Ir/Ru-Ir/Ru bond lengths at room temperature as a function of $x$ for \liro. The inset of (a) schematically illustrates the regular triangular lattice of the trigonal phase and the isosceles triangular lattice of the monoclinic phase. The thick solid lines represent short chemical bonds and the thin solid lines represent long chemical bonds.}
\end{figure}

At room temperature, \lio\ and \lro\ have the monoclinic symmetry with space group $C2/m$ and $P2_{1}/m$, respectively  \cite{JPSJ.76.033705}. In order to determine where the structure is changed from $P2_{1}/m$ to $C2/m$ with increasing $x$, the XRD patterns for whole series are tried to be fitted by both crystallographic structures. We found \cite{note:SI} that when $x$ is closed to 1, the patterns can only be fitted well by using the space group of $P2_{1}/m$, on the other hand, the space group of $C2/m$ has better fitting quality than $P2_{1}/m$ near \lio\ side. But in between, especially when $x$ is near 0.5, the patterns can be fitted by using either of structural models. From the values of the fit residuals ($R_{p}$ and $R_{wp}$), the crossover from $P2_{1}/m$ to $C2/m$ happens at $x=0.5-0.6$. The fitted lattice parameters, the unit cell volume $V_{cell}$ and bond lengths between Ir/Ru and Ir/Ru as a function of $x$ at room temperature is shown in Fig.~1. The $a$- and $b$-axial lattice parameters decrease with increasing $x$ for both crystallographic structures; on the other hand, the $c$-axial lattice parameter slightly increases in general for the $C2/m$ space group and remains almost unchanged for the $P2_{1}/m$ space group [Fig.~1(a)]. For the values of the $\beta$ angle, both of them decrease with $x$ [Fig.~1(b)]. Fig.~1(c) shows the change of volume of unit cell as a function of $x$. It can be seen that the unit cell shrinks gradually with Ru doping, and the obtained values for the two structures are consistent. It can be ascribed to the slightly smaller ionic radius of Ru$^{4+}$ (0.67 \r{A}) than Ir$^{4+}$ (0.68 \r{A}).

The most important structural parameters are the bond lengths between Ir/Ru and Ir/Ru ions. Fig.~1 (d) shows the existence of two $x$ regimes:
With Ru doping, the shortest Ru-Ru bond length (Ru/Ir1-Ru/Ir4) decreases gradually for $x > 0.5$, before which that bond length is insensitive with $x$. On the other hand, the other two Ir/Ru-Ir/Ru bond lengths (Ir/Ru1-Ir/Ru2 and Ir/Ru1-Ir/Ru3) are nearly unchanged and the difference between these two bond lengths are within $3.3\%$ for the whole series, compared with the $14\%$ difference between Ir/Ru1-Ir/Ru2 and Ir/Ru1-Ir/Ru4 in \lro.

\begin{figure}[tbp]
\centerline{\includegraphics[scale=0.18]{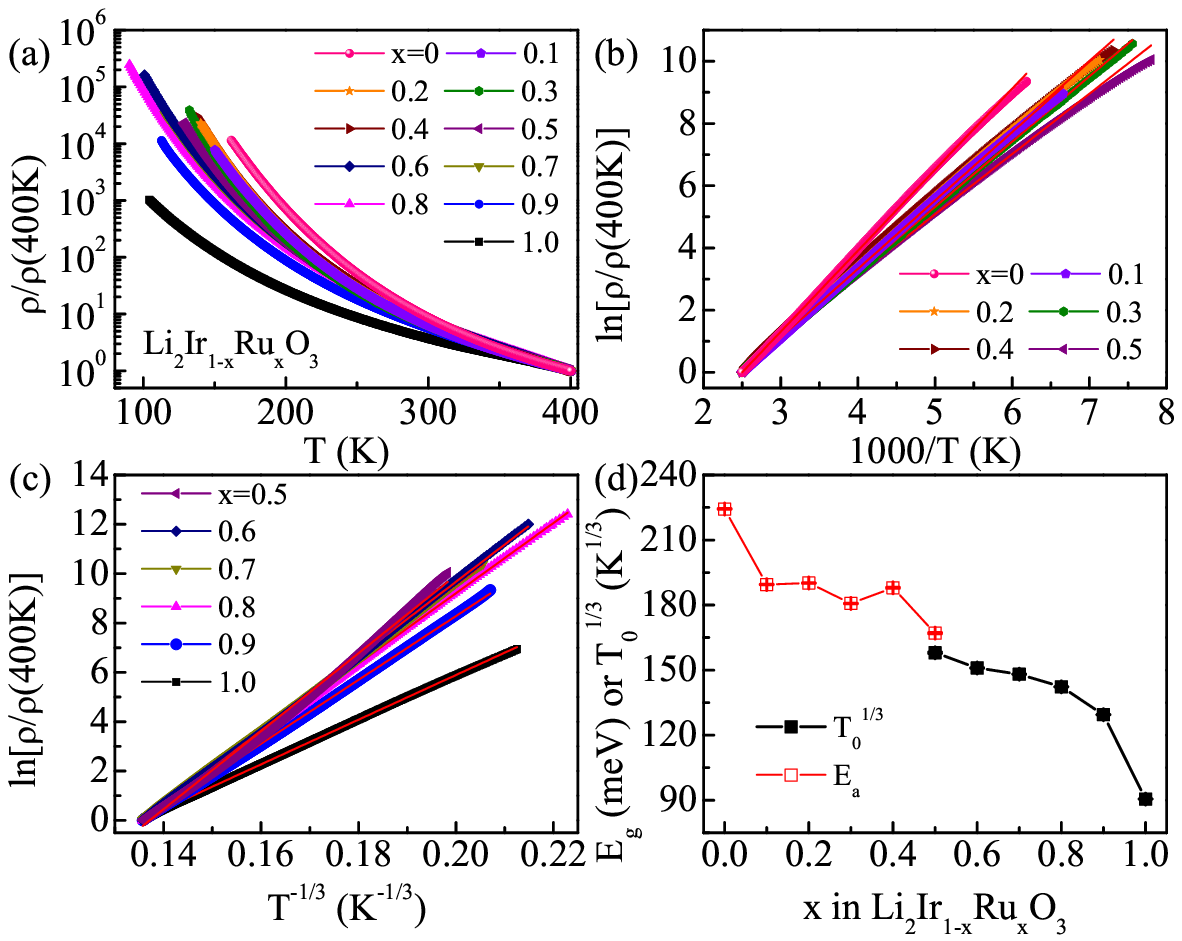}} \vspace*{-0.3cm}
\caption{(a) Temperature dependence of reduced resistivity $\rho(T)/\rho($400K$)$ of the \liro\ polycrystal. The $\rho($400K$)$ from $x=$ 0 to $x=$ 1 is 27.7, 18.1, 11.2, 8.6, 9.5, 7.8, 2.9, 4.4, 1.7, 7.9, and 8.3 $ \Omega$ cm, respectively. The fits of $\rho(T)/\rho($400K$)$ curves using (b) thermal activation model for $x\leqslant$ 0.5 and (c) variable-range hopping model for 0.5 $\leqslant x\leqslant$ 1.0. (d) Fitted thermal activation energy and the characteristic temperature $T_{0}^{1/3}$.}
\end{figure}

Regarding the electron transport properties, all of the temperature dependencies of resistivity $\rho $(T) for the whole series show insulating behaviors [Fig.~2(a)]. In particular, the resistivity decreases monotonically as $x$ increases, in sharp contrast with the decrease-then-increase behavior generally expected for charge doping between two insulating end members.
The insulating behaviors for both ending members \lro\ and \lio\ below 400 K are consistent with previous experimental results in the literature \cite{Na2IrO3:Singh,JPSJ.76.033705}. Quantitatively, the insulating behavior for $x\leq$ 0.5 follows the Arrhenius law $\rho=\rho_{0}\exp(E_{a}/T)$ very well, where $E_{a}$ is thermal activation energy [Fig.~2(b)]. It is different from that of Na$_{2}$IrO$_{3}$ where the three-dimensional variable-range hopping (VRH) mechanism seems dominate the resistivity behavior \cite{PhysRevB.82.064412}. However, when $x\geq$ 0.5, the behaviors of $\rho(T)$ start to deviate from Arrhenius law and crossover to VRH region ($\rho=\rho_{0}\exp[(T_{0}/T)^{1/(d+1)}]$, where $T_{0}$ is the characteristic temperature and $d$ is the dimension of system) [Fig.~2(c)]. The fitting results for $d=$ 2 is slightly better than that for $d=$ 3, implying that the dimensionality of VRH might be two-dimensional. As shown in Fig.~2(d), the fitted $E_{a}$ and $T_{0}^{1/3}$ decreases with Ru doping generally.

\begin{figure}[tbp]
\centerline{\includegraphics[scale=0.3]{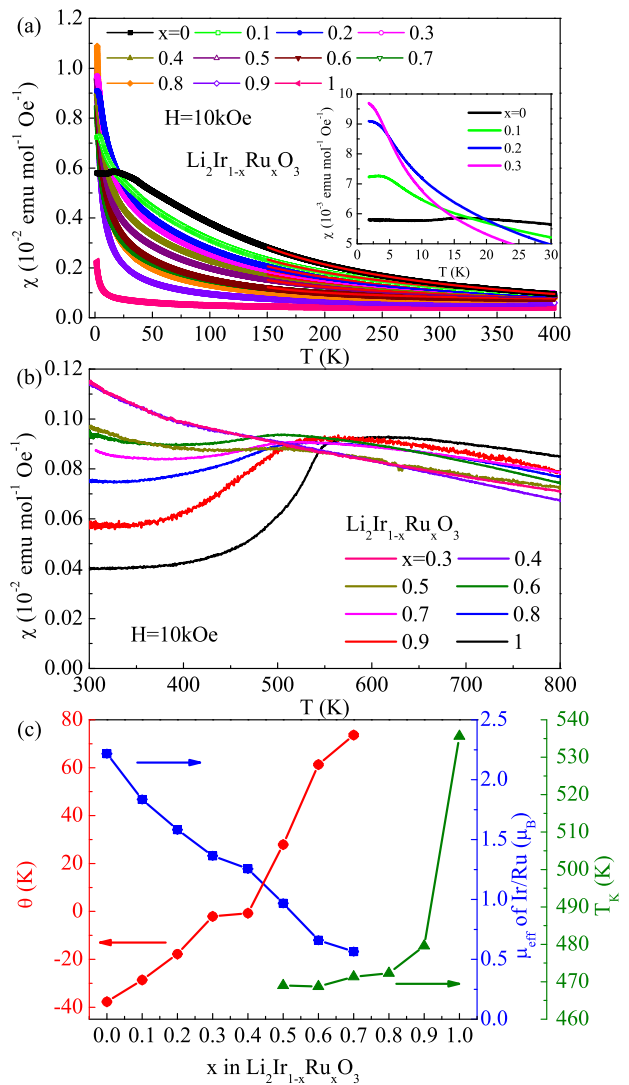}} \vspace*{-0.3cm}
\caption{(a) Temperature dependence of dc magnetic susceptibility $\chi(T)$ of the \liro\ polycrystal between 2 - 400 K at H = 10 kOe with Zero Field Cooling (ZFC) mode. Inset: enlarged part of $\chi(T)$ at low temperature for 0.7 $\leqslant x\leqslant$ 1. (b) Temperature dependence of $\chi(T)$ between 300 - 800 K at H = 10 kOe for 0 $\leqslant x\leqslant$ 0.7. (c) Fitted Weiss temperature $\theta$, effective moment $\mu_{eff}$ of Ir/Ru and transition temperature $T_{K}$ at high temperature as function of $x$ for \liro.}
\end{figure}

The magnetic susceptibility as a function of temperature $\chi(T)$ is shown in Fig.~3(a). For Li$_{2}$IrO$_{3}$, there is a drop at around 15 K, which is consistent with the result in literature and ascribed to the AFM transition \cite{Na2IrO3:Singh}. Fitting the $\chi(T)$ data between $T=$ 150 and 400 K using the Curie-Weiss law
\begin{equation}
\chi(T)=\chi_{0}+C/(T-\theta),
\end{equation}
we obtained that $\theta=-37.7(2)$ K and $C=0.6174(9)$ emu K/mol Oe. The Weiss temperature $\theta$ is close to the reported value previously \cite{Na2IrO3:Singh}. Assuming the $g$ factor equals 2, the obtained value of $C$ corresponds to an effective moment of $\mu_{eff}=$ 2.217(2) $\mu_{B}$ for Li$_{2}$IrO$_{3}$, which is slightly larger than 1.83(5) $\mu_{B}$ reported in Ref. \onlinecite{Na2IrO3:Singh}. This value suggests that the spin moment of Ir$^{4+}$ is 1/2. The frustration factor $f=|\theta|/T_{N}\approx 2.48$.
With increasing the content of Ru, the AFM transition is suppressed quickly [inset of Fig.~3(a)]. When $x=0.1$, $T_{N}$ is shifted to about 3~K and $\theta=-28.6(1)$~K; thus the nominal value of $f$ increases to about 9.5. With further increasing of Ru, the AFM transition becomes incomplete for $x=0.2$ and cannot be observed for $x>0.3$ down to $2$~K.
It is tempting to attribute the enhancement of magnetic frustration and the suppression of the AFM order to the emergence of the KSL state in the way that Ru doping promotes relatively the Kitaev interaction to dominate the Heisenberg exchange interaction. For the lightly doped regime, this scenario seems compatible with the observed insulating character, as
no quasiparticles were found in the KSL regime in recent studies of a single hole moving in the Kitaev-Heisenberg model \cite{arXiv:1308.3373,PhysRevLett.111.037205}. Yet, it is unlikely to apply to larger $x$. As shown in Fig.~3(c), the increase of Ru content changes the sign of the Weiss temperature $\theta$ from negative to positive around $x=0.4$, while the fitted effective moment of Ir/Ru keeps getting smaller. This signals that a different mechanism starts taking over the low-energy physics.

This shift of physics becomes apparent in the other end member \lro. As shown in Fig.~3(b), the $\chi(T)$ curve for $x=1$ drops to a very small value around $T_K=540$ K, leading to the nearly temperature-independent behavior [$\chi_{0}$ in Eq.~(1)] below 400 K. This is consistent with the dimerization of the Ru-Ru bonds \cite{JPSJ.76.033705,JPSJ.78.094706,PhysRevLett.100.147203,arXiv:1310.7810}. Upon reducing the amount of Ru, the transition temperature $T_{K}$ shifts to a lower temperature and the changes of $\chi(T)$ at $T_{K}$ become smaller [Figs.~3(b) and 3(c)]. Finally, this anomaly cannot be observed at $x \sim 0.5$. These results are consistent with the above crystallographic data where the shortest Ir/Ru-Ir/Ru bond length increases gradually with decreasing $x$ and it becomes comparable with the other two bonds at $x \sim 0.5$. Furthermore, we found $\chi_{0}(x) \approx \alpha x$ where $\alpha=0.00111(9)$ emu/mol-Oe for $0.1<x<0.8$ \cite{note:SI}. This means that the portion of the bond dimers increases with Ru doping and could exist (fluctuate) well below $x=0.5$, generating electronic inhomogeneity.

\begin{figure}[b]
\centerline{\includegraphics[width=\columnwidth,clip=true,angle=0]{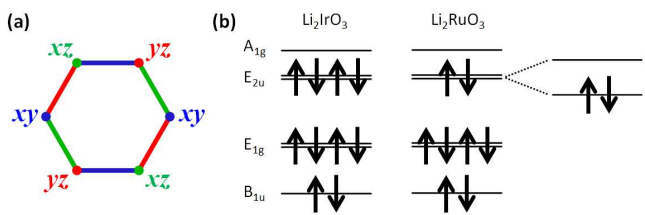}} \vspace*{-0.3cm}
\caption{Schematic plots of (a) a (Ir,Ru)$_6$ hexagon with six relevant $t_{2g}$ orbitals in the tight-binding approximation and (b) the energy levels of the six quasi-molecular orbitals formed by the six $t_{2g}$ atomic orbitals on one hexagon~\cite{Na2IrO3:Mazin} with electron filling corresponding to \lio\ (left) and \lro\ (right), rendering \lro\ subject to the Jahn-Teller splitting of the half-filled doubly degenerate $E_{2u}$ level.}
\end{figure}

Since Ru$^{4+}$ $4d^4$ is a strong impurity scattering center to Ir$^{4+}$ $5d^5$, it is reasonable to ask whether the Anderson localization is at work in \liro. In a comparative study, we found similar structural, magnetic, and electric behaviors in \lrro\ \cite{note:lirho}, where Ru$^{4+}$ is a weak impurity scattering center to Rh$^{4+}$ $4d^5$, since they are nearest neighbors in the periodic table. The Ir/Ru (or Rh/Ru) disorder is thus unlikely the driving force for the persistent insulating behavior in these 213 systems, which is supported by the fact that the Ru substitution for Ir can drive a robust metallic state in \siro\ ($x=0.5$) \cite{PhysRevB.86.125105}. Interestingly, there could exist an ``intrinsic'' source of strong disorder in A$_2$IrO$_3$, namely, the A sites centered at the hexagons of the Ir sublattice could be partially occupied by Ir and \emph{vice versa} \cite{PhysRevB.82.064412,hi_Shikano_Kageyama_Kanno_2003}. It remains to be elucidated how the degree of the Li/Ir disorder is affected by the Ru substitution.

We emphasize that the strong bond dimerization points to the large effect of EP coupling, which was long recognized to be critical for the persistent insulating character of the Mott insulators \cite{Mott}. The different transport behaviors exhibited by the doped 213 and 214 iridates are reminiscent of the historic comparison of the doped nickelate La$_{2-x}$Sr$_x$NiO$_4$ that remains insulating and the doped cuprate La$_{2-x}$Sr$_x$CuO$_4$ that becomes metallic for $x>0.03$ \cite{PhysRevB.43.1229,PhysRevLett.71.2461,PhysRevLett.111.096404}. In the nickelate both $3d_{z^2}$ and $3d_{x^2-y^2}$ orbitals are active, while in the cuprate only the latter one is. The orbital d.o.f. interplaying with the charge and spin d.o.f. generally results in a synergy between electron-electron and EP interactions which reinforce each other to drive a stronger tendency to small polarons, domain walls, and charge-density waves in the nickelates than in the cuprates \cite{PhysRevB.50.7222,PhysRevLett.70.2625}. Likewise, charge ordering and large effects of EP coupling were often seen in the Jahn-Teller active manganites with degenerate $3d_{z^2}$ and $3d_{x^2-y^2}$ orbitals \cite{CO:manganite,PhysRevLett.96.116405}. Here we argue that the essential physics underlying the persistent insulating character of \liro\ is similar---the doped holes experience the orbital d.o.f. and large effects of EP coupling---rather than the common one-band $j_\mathrm{eff}=1/2$ modeling with EP interaction neglected.

In the following, we describe a possible origin of the hole induced breakdown of the $j_\mathrm{eff}=1/2$ picture  employing the quasi-molecular-orbital (QMO) concept recently proposed for A$_2$IrO$_3$ \cite{Na2IrO3:Mazin}. As illustrated in Fig.~4(a), a peculiar feature of the Ir honeycomb lattice is that although every Ir site contributes three $5d$ $t_{2g}$ ($xy$, $yz$, $zx$) orbitals to the low-energy physics, only one $t_{2g}$ orbital is relevant to a given Ir$_6$ hexagon in the tight-binding approximation. As a result, the hexagons could be approximately treated as independent building blocks of the lattice and the energy levels are determined by forming six molecular orbitals per hexagon \cite{Na2IrO3:Mazin}. The electron filling is ten electrons for \lio\ and eight for \lro\ [Fig.~4(b)]. In both cases, the highest occupied QMOs are two-fold degenerate with the $E_{2u}$ symmetry. They are fully and half occupied in \lio\ and \lro, respectively. This leads to the Jahn-Teller instability in \lro\ where the strong bond dimerization is now viewed as a novel QMO Jahn-Teller distortion, but in \lio\ such instability is absent and the $j_\mathrm{eff}=1/2$ local state can be stabilized. This local-hexagon picture provides a simple explanation of the bond dimerization and its persistence in a wide range of doping levels. It also agrees with recent local structural x-ray measurements showing that disordered dimers survive at the nanoscale up to at least $920$~K \cite{arXiv:1310.7810}. An intriguing implication of this picture is inhomogeneous deformation of the hexagons with charge disproportion among them in hole-doped A$_2$IrO$_3$ systems including A$_{2-x}$IrO$_3$.

In summary, we have studied the structural, magnetic, and electric properties of \liro\ polycrystals. We found that this system remains insulating for all the doping levels, contrary to the predictions based on the widely used $j_\mathrm{eff}=1/2$ Kitaev-Heisenberg model. Our analyses suggest that the hole-doped honeycomb iridate system be a unique $5d$-orbital-based platform to study the interplay of the charge, orbital, spin, and lattice degrees of freedom, which warrants further investigation. 


This work was supported by the Funding Program for World-Leading Innovative R\&D on Science and Technology (FIRST), Japan. The work at Brookhaven National Laboratory was supported by the U.S. Department of Energy (DOE), Division of Materials Science, under Contract No. DE-AC02-98CH10886.

\end{document}